# BeyondCT: A deep learning model for predicting pulmonary function from chest CT scans

Kaiwen Geng[1], Zhiyi Shi[1], Xiaoyan Zhao[1], Alaa Ali[1], Jing Wang[1], Joseph Leader[1], Jiantao Pu[1, 2]

[1] Department of Radiology, University of Pittsburgh, Pittsburgh, PA 15213, USA

[2] Department of Bioengineering, University of Pittsburgh, Pittsburgh, PA 15213, USA

**\*Corresponding authors and guarantors of the entire manuscript:**

Jiantao Pu, PhD

Contact Phone: (412) 641-2571, email: jip13@pitt.edu

**Abstract**

**Background**: Pulmonary function tests (PFTs) and computed tomography (CT) imaging play crucial and complementary roles in clinical practice for diagnosing, managing, and monitoring various lung diseases. In practical settings, a common problem is the absence of or ready access to recorded pulmonary functions, despite the existence of chest CT scans.

**Purpose**: To develop and validate a deep learning algorithm to predict pulmonary function directly from chest CT scans.

**Methods**: The development study cohort originated from the Pittsburgh Lung Screening Study PLuSS) (n=3619). The validation cohort originated from the Specialized Centers of Clinically Oriented Research (SCCOR) in chronic obstructive pulmonary disease (COPD) (n=662). A deep learning model called *BeyondCT* combined a three-dimensional (3D) convolutional neural network (CNN) and the Vision Transformer (ViT) architecture was used to predict forced vital capacity (FVC) and forced expiratory volume in one second (FEV1) from non-contrasted inspiratory chest CT scans. A 3D CNN model without ViT was used for comparison, Subject demographics (e.g., age, gender, smoking status) were also incorporated into the model. Model performance was compared to actual PFT measurements using mean absolute error (MAE, unit: L), percentage error, and R square ($R^2$).

**Results**: The 3D-CNN model achieved MAEs of 0.395 L and 0.383 L, percentage errors of 13.84% and 18.85%, and $R^2$ of 0.665 and 0.679 for computing FVC and FEV1, respectively. The BeyondCT model without demographics achieved MAEs of 0.362 and 0.371, percentage errors of 10.89% and 14.96%, and $R^2$ of 0.719 and 0.727 for computing FVC and FEV1, respectively. When demographics were included, The BeyondCT model with demographics demonstrated a significant improvement ($p<0.05$), yielding an MAE of 0.356 and 0.353, percentage errors of 10.79% and 14.82%, and $R^2$ of 0.77 and 0.739 for FVC and FEV1 in the independent test set, respectively.

**Conclusion**: The BeyondCT model demonstrated a relatively accurate and robust performance for predicting lung function from non-contrast inspiratory chest CT scans.

# I. Introduction

Pulmonary function tests (PFTs) and chest computed tomography (CT) imaging are widely utilized to assess lung disease, such as lung cancer, asthma, chronic obstructive pulmonary disease (COPD), and pulmonary fibrosis. PFTs are used to measure lung capacity, airflow, and gas exchange to assess how well the lungs are functioning. CT scans offer detailed anatomical information about the lungs and can detect structural abnormalities, such as tumors, emphysema, or fibrosis, as well as their changes over time. These two tests are often used together to complement each other's diagnostic capabilities. Their combination can provide a more comprehensive assessment of lung disease and allow clinicians to better understand the underlying pathology and determine the appropriate treatment. However, in practical settings, pulmonary function measures are not always readily accessible along with chest CT scans. While spirometry based PFTs are relatively simple and can be conducted in various healthcare settings, they may still require additional effort and scheduled appointments. Moreover, PFTs may yield inaccurate results when patients are not in optimal physical condition. Consequently, structural anomalies observed in CT scans without functional data can only provide a partial view. If there is a tool available that can infer pulmonary function measures directly from chest CT scans and/or patient demographics, it will provide clinicians with a convenient and efficient means to gather essential functional information, enabling more accurate and efficient diagnoses and tailored treatment plans.

Several studies explored the feasibility of inferring lung function measures from images and demographics. Chen et al. [1] used a multi-output support vector regression (SVR) model to predict FVC and FEV1 based on demographic and inflammatory parameters. This model was proposed to facilitate the diagnosis and treatment of lung disease when patients could not undertake a pulmonary function test, such as during acute exacerbations of COPD. Their model showed an overall MSE of 0.17 and an $R^2$ of 0.39 in FVC prediction and achieved an MSE of 0.19 and $R^2$ of 0.29 in FEV1 prediction. Wang et al. [2] developed a model to predict FVC for patients who were unable to meet the complete exhalation requirement of spirometry due to health constraints. They utilized age, gender, FEV1, and Peak Expiratory Flow (PEF) as input features from 354 subjects for an SVR model. Their model demonstrated a sensitivity of 92% in identifying the subjects with abnormal lung functions.

Recently, deep learning (DL) techniques, such as three-dimensional (3D) convolutional neural network (CNN), have been employed to infer pulmonary function from various clinical images. Park et al. [3] described a DL algorithm to predict FVC and FEV1 from low-dose CT scans. Their model, trained on 16,148 participants who underwent both CT scan and spirometry, demonstrated strong agreement between CT derived metrics and spirometry (concordance correlation coefficient (CCC): FVC = 0.94, FEV1 = 0.91), low bias (MAE: FVC = 0.22L, FEV1 = 0.22L), and achieved an accuracy of 90.2% in classifying COPD high-risk individuals using FEV1/FVC < 0.7 as the threshold. While this approach showed promise, it exhibited a wide range of sensitivities (36.1%-61.6%), indicating potential limitations in identifying all high-risk cases and potential overfitting. Javaregowda et al. [4] proposed a CNN-based approach for assessing the progression of Idiopathic Pulmonary Fibrosis (IPF) by forecasting FVC values over a span of 2.5 to 3 years. Their model incorporated lung CT scans, initial FVC measurement, age, gender, and smoking status as the inputs and achieved a mean percentage error of 1.23% on FVC prediction. Schroeder et al. [5] developed a CNN-based model to predict FEV1/FVC using 6,749 2-view chest radiographs. The model achieved an area under the receiver operating characteristic (ROC) curve (AUC) of 0.814. Their CNN-based model consistently outperformed natural language processing baseline models trained on radiologist text reports (AUC = 0.770) in predicting severe COPD (p-value < 0.05).

We present a novel DL model called BeyondCT to predict FVC and FEV1 from non-contrast, inspiratory chest CT scans and demographics. This model combined 3D CNN with the Vision Transformer (ViT) architecture [6, 7]. The 3D CNN module is responsible for extracting features from augmented input data, while the ViT module is used to generate predictions from the extracted feature map. Additionally, we incorporated demographics through an optional multimodal integration. Two cohorts were utilized to develop and validate the BeyondCT model. The low-dose chest CT (LDCT) scans acquired from a lung cancer screening program were used to develop the DL model, while the diagnostic CT scans acquired from a COPD study cohort were used to validate the DL model independently. A detailed description of the BeyondCT model and its validations follows.

## II.    Materials and Methods

### A. Study cohorts

(1) Algorithm development cohort: Pittsburgh Lung Screening Study (PLuSS). This cohort consists of 3,642 subjects enrolled in an LDCT-based lung cancer screening program. These subjects were enrolled between 2002 and 2005. Inclusion criteria: 50-79 years old, and current or ex-cigarette smokers with at least 12.5 pack-years at time of enrollment. Exclusion criteria: 1) quit smoking >10 years earlier, 2) reported a history of lung cancer, or 3) reported chest CT within one year of enrollment. Demographic and smoking history data was collected using structured interviews and questionnaires at baseline, and smoking status is updated yearly. Participants underwent spirometry and LDCT screening within 2 weeks. The lung cancer screening protocol used a single-breath-hold, helical, low-dose technique (40–60 mAs, 120 kVp). Images were reconstructed with a high spatial frequency (lung) algorithm at 2.5 mm thickness and spacing. A total of 11,096 LDCT scans acquired on 3,619 subjects with spirometry data were used in this study after removing CT scans with poor image quality (Table 1).

**Table 1.** PLuSS demographics (n=3,619)

| Characteristics | | Overall (n = 3,619) | Train (n = 2,895) | Validation (n = 362) | Test (n = 362) |
|---|---|---|---|---|---|
| Age, years (±SD) | | 60.04 ±12.99 | 60.02 ±12.93 | 59.96 ±13.87 | 60.23 ±12.59 |
| Sex, n (%) | Male | 2,059 (56.9%) | 1,644 (56.8%) | 207 (57.2%) | 212 (58.5%) |
| | Female | 1,560 (43.1%) | 1,251 (43.2%) | 155 (42.8%) | 150 (41.5%) |
| Cigarettes Per Day, n (±SD) | | 25.26 ±8.63 | 25.21 ±8.61 | 25.44 ±8.74 | 25.52 ±8.68 |
| Height, inches (±SD) | | 66.7 ±3.71 | 66.72 ±3.71 | 66.51 ±3.72 | 66.76 ±3.73 |
| Weight, lbs (±SD) | | 182.36 ±39.95 | 182.44 ±40.1 | 180.77 ±37.79 | 183.31 ±40.83 |
| Smoke Duration, years (±SD) | | 41.04 ±7.59 | 41.06 ±7.6 | 41.18 ±7.33 | 40.81 ±7.72 |
| Emphysema, n (%) | Yes | 1,770 (48.9%) | 1,418 (49.0%) | 175 (48.5%) | 178 (49.2%) |
| | No | 1,849 (51.1%) | 1,477 (51.0%) | 187 (51.5%) | 184 (50.8%) |
| GOLD Criteria, n (%) | 0 | 1,657 (45.8%) | 1327 (45.8%) | 167 (46.1%) | 163 (45.0%) |
| | 1 | 684 (18.9%) | 543 (18.8%) | 64 (17.7%) | 77 (21.3%) |
| | 2 | 916 (25.3%) | 738 (25.5%) | 89 (24.6%) | 89 (24.6%) |
| | 3 | 322 (8.9%) | 256 (8.8%) | 37 (10.2%) | 29 (8.0%) |
| | 4 | 40 (1.1%) | 31 (1.1%) | 5 (1.4%) | 4 (1.1%) |
| Smoking Status, n (%) | Current | 2,229 (61.6%) | 1779 (61.5%) | 226 (62.4%) | 224 (61.9%) |

| | Former | 1,390 (38.4%) | 1116 (38.5%) | 136 (37.6%) | 138 (38.1%) |
| --- | --- | --- | --- | --- | --- |
| **FVC, liters (±SD)** | | 3.36±1.03 | 3.36±1.01 | 3.33±0.99 | 3.33±0.98 |
| **FEV1, liters (±SD)** | | 2.27±0.85 | 2.28±0.86 | 2.21±0.79 | 2.22±0.84 |

We randomly divided the subjects in the PLuSS cohort into the training set (n = 2,895, mean age: 60.02 years ± 12.93 (SD); male: 56.8%, CT scans: 8,877), validation set (n = 362, mean age, 59.96 years ± 13.87 (SD); male: 57.2%, CT scans: 1,106), and an independent test set (n = 362, mean age, 60.23 years ± 12.59 (SD); male: 58.5%, CT scans: 1,113).

**(2) Independent test cohort: Specialized Centers of Clinically Oriented Research (SCCOR).** This cohort consists of 2,287 CT scans acquired on 662 subjects. These subjects were either current or former smokers with a history of consuming at least 10 packs per year and 50 years or older. The cases have comprehensive records of PFTs, chest CT scans, and demographics (Table 2). Pre- and post-bronchodilator body plethysmography was performed in SCCOR with post-bronchodilator values used in this study. Three consistent maneuvers were required for both FVC and FEV1. Subjects were classified into Global Initiative for Chronic Obstructive Lung Disease (GOLD) five categories: non-COPD (256 subjects), GOLD-I (91 subjects), GOLD-II (157 subjects), GOLD-III (71 subjects), and GOLD-IV (87 subjects).

The CT scans were acquired using a protocol with a single-breath-hold at end inspiration without contrast at the following parameters: 64×0.625 mm detector configuration, 0.969 pitch,120 kVp tube energy, 250 mA tube current, and 0.4 sec gantry rotation (or 100 mAs). Images were reconstructed to encompass the entire lung field in 512×512 pixel matrix using the GE "bone" kernel at 0.625 mm section thickness and 0.625 mm interval. Pixel dimensions ranged from 0.549 to 0.738 mm, depending on the participant's body size.

**Table 2**. SCCOR demographics (n=662)

| **Characteristics** | | **Count (n = 662)** |
| --- | --- | --- |
| **Age, years (±SD)** | | 65.01±5.19 |
| **Sex, n (%)** | Male | 362 (54.7%) |
| | Female | 300 (45.3%) |
| **Cigarettes Per Day, n (±SD)** | | 29.04±16.39 |
| **Height, inches (±SD)** | | 66.62±3.79 |
| **Weight, lbs (±SD)** | | 173.38 ±35.17 |
| **Smoking Status, n (%)** | Current | 256 (38.7%) |
| | Former | 406 (61.3%) |
| **FVC, liters (±SD)** | | 3.38 ±0.95 |
| **FEV1, liters (±SD)** | | 2.11 ±085 |

The cohorts were acquired under the University of Pittsburgh Institutional Review Board (IRB) approved protocols (PLuSS: #21020128, SCORR: #0612016), and written informed consent was obtained from each subject.

## B. BeyondCT architecture

The BeyondCT model is a fusion of 3D CNN and ViT models (Fig.1). To ensure consistent inputs, the CT scans were isotropicized with a resolution of $1.5 \times 1.5 \times 1.5$ mm$^3$ and padded with a uniform dimension of $256 \times 256 \times 256$ voxels. The one-channel CT pixel input data were initially processed by a 3D CNN network consisting of three convolutional layers with a stride of [2, 2, 2] and a kernel of $2 \times 2 \times 2$. This CNN-based network transformed the input CT data into feature maps of size $32 \times 32 \times 32$ with 8 channels. Subsequently, the feature maps were partitioned into $4 \times 4 \times 4$ image patches with 8 channels, and these patches were flattened into one-dimensional vectors of size 512 and passed through a linear layer for patch embedding. Optionally, demographics could be embedded into a patch using another linear layer. The embedded image patches, embedded demographics patches, and positional information were then combined and fed into 4 transformer blocks with 8 heads each.

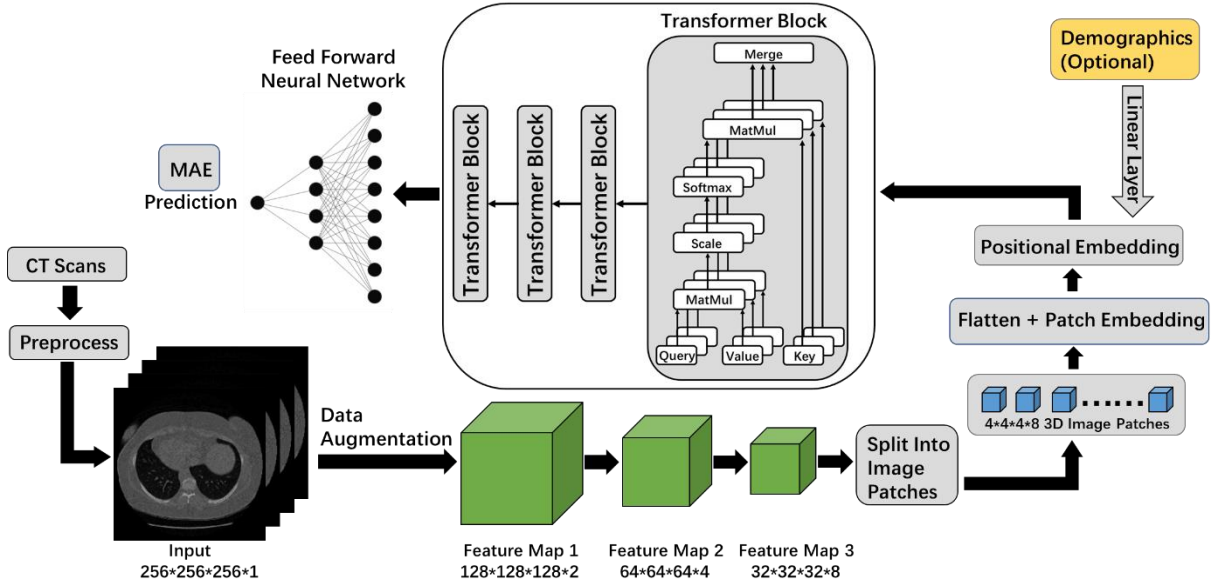

**Figure 1:** BeyondCT Architecture. The initial module includes a 3D CNN that extracts feature maps from CT scans using three convolutional layers. These feature maps are subsequently partitioned into $4 \times 4 \times 4 \times 8$ patches and flattened into 1D vectors using patch embedding. Patient demographics can be optionally incorporated. The embedded image patches, along with the optional embedded demographics patches, are then processed by a ViT consisting of 4 blocks and 8 heads. Mean absolute errors are computed by comparing the predicted FVC/FEV1 with the actual FVC/FEV1 values and are used as the loss function.

The core part of the transformer blocks is the multi-head self-attention mechanism [8]. In the $i^{th}$ attention layer, given the input tensor $Input \in \mathbb{R}^{Height \times Weight \times Channels_{in}}$ , the corresponding output can be computed by:

$$Output^{(i)} = Attention(Q^{(i)}, K^{(i)}, V^{(i)}) = softmax\left(\frac{Q^{(i)}K^{(i)T}}{\sqrt{d_k}}\right)V^{(i)} \qquad (1)$$

$$Output^{(i)} \in \mathbb{R}^{Height \times Weight \times Channels_{output}}$$

where $Q^{(i)}$ (Queries) $= Input \times W_Q^{(i)}$ $K^{(i)}$(Keys) $= Input \times W_K^{(i)}$ $V$ (values) $= Input \times W_V^{(i)}$. $W_Q^{(i)}, W_K^{(i)}, W_V^{(i)} \in \mathbb{R}^{Channels_{in} * Channels_{output}}$ are learnable parameters at attention layer i and $d_k$ is the scaling factor indicating the dimensionality of the Key vectors and is the same across all the attention layers. Multi-head attention in Transformers enables the model to capture diverse information from distinct representation subspaces concurrently. Each attention head focused on different aspects of the input, facilitating a more comprehensive understanding and quantification of the interactions of each patch and all other patches in the sequence.

Following this, three linear layers are utilized to generate predictions for either FVC or FEV1. The mean absolute error (MAE) between the predicted and the actual value was used as the loss function.

Demographics were incorporated into the model using a numerical vector containing information such as age, sex (female: 0, male:1), height, body weight, smoking status (current: 0, former: 1), and smoking history were passed through a linear layer without applying normalization to adjust its dimension so that it could match the dimension of the embedded image patches [9]. The demographic information vector was then merged with the patch-embedded image vector through concatenation. The concatenated results were combined with positional information and subsequently fed into the transformer blocks for further processing.

**C. Model Training:**

To enhance the robustness of our model, data augmentation was performed on-the-fly during the model training. The data augmentation encompassed a range of transformations to increase the diversity of CT data, including voxel value shift (altering voxel values by -25% to 25%), contrast scaling (within a factor of 0.8 to 1.2), random cropping or padding (within 10% of the original size), horizontal or vertical flipping (50% probability), and independent scale transformations on the x and y-axes (ranging from 80% to 120%). Additionally, we introduced random image translation within a margin of 15% of the original size in both the x and y directions, rotation within the range of -90 to 90 degrees, and shear transformation within a range of -15 to 15 degrees. To address data heterogeneity, we applied blur effects using methods, such as Gaussian blur, average blur, or median blur, and introduced Gaussian noise into the image data, with the standard deviation of the noise randomized between 0 and 51. Each CT scan in the training set underwent a random subset of these transformations.

Due to limited computational memory, the batch size was set at 2. The Adam optimizer was employed for model optimization and parameter updates as it demonstrated high convergence speed [10]. The epoch number was set at 100. After each training epoch, the model's accuracy on the internal validation set was computed and the model with the highest accuracy was saved. The entire training procedure was performed on an NVIDIA GeForce RTX 3060 GPU, utilizing the PyTorch 1.12.1 framework.

**D. Performance Evaluation**

Mean absolute error (MAE, | Actual − Predicted | ,unit: liter), percentage error (% error), and R squared ($R^2$) were used as the performance metrics to evaluate the performance of the model for predicting FVC and FEV1. For comparison purposes, we implemented the 3D CNN model described by Park et al. [3] and a linear regression model based on sole patient demographics. All prediction models were evaluated on the independent test cohort. The impact of the inclusion of patient demographics on the performance of the BeyondCT model was also evaluated. The paired sample t-test was used to compare the performance differences between two models. A p-value less than 0.05 was considered statistically significant.

Additionally, subjects in the PLuSS independent test cohort were grouped by gender, presence of emphysema, smoking status, and GOLD criteria. Evaluating BeyondCT's performance on these subgroups aimed to ascertain the consistency of its predictive capabilities across these different criteria.

## III. Results

**A. Performance on the PLuSS cohort**

The performance of the developed models for predicting FVC and FEV1 on the PLuSS cohort was summarized in Table 3. The BeyondCT model with the CT data and demographics had the best performance predicting both FEV1 and FVC in the PLuSS independent test set (Table 1 and Fig. 2). The model had MAEs of 0.356 L and 0.353 L, percentage errors of 10.8% and 14.8%, and R²s of 0.770 and 0.739 for predicting FVC and FEV1, respectively. The BeyondCT with demographics model performance metrics were significantly better than all other models (p < 0.05). The BeyondCT model with demographics had mean differences of -0.01 L and -0.02 L, lower limits of agreement of -2.1 L and -1.8 L, and upper limits of agreement of 2.0 L and 1.9 L for FVC and FEV1, respectively, in the Bland Altman analysis (Fig. 3). The difference between the observed and predicted values for both FVC and FEV1 were equally distributed and unbiased across the mean FVC and FEV1 values (Fig.3).

**Table 3**: Models to predict FVC and FEV1 in the PLuSS independent test set (n=362).

| | | Linear regression with demographics | | | 3D CNN Baseline | | | BeyondCT with demographics | | | BeyondCT without demographics | | |
|---|---|---|---|---|---|---|---|---|---|---|---|---|---|
| | | MAE (L) | %Error | $R^2$ | MAE (L) | %Error | $R^2$ | MAE (L) | %Error | $R^2$ | MAE (L) | %Error | $R^2$ |
| **FVC** | Train | 0.514 | 16.98% | 0.553 | 0.376 | 12.41% | 0.692 | 0.359 | 10.17% | 0.724 | 0.351 | 10.03% | 0.773 |
| | Val | 0.516 | 16.79% | 0.507 | 0.387 | 13.79% | 0.668 | 0.363 | 10.93% | 0.728 | 0.357 | 10.84% | 0.762 |
| | Test | 0.515 | 16.86% | 0.511 | 0.395 | 13.84% | 0.665 | 0.362 | 10.89% | 0.719 | 0.356 | 10.79% | 0.77 |
| **FEV1** | Train | 0.486 | 28.19% | 0.377 | 0.371 | 17.21% | 0.693 | 0.367 | 14.04% | 0.732 | 0.352 | 13.95% | 0.744 |
| | Val | 0.498 | 27.65% | 0.323 | 0.385 | 18.74% | 0.684 | 0.368 | 14.79% | 0.718 | 0.355 | 14.68% | 0.741 |
| | Test | 0.491 | 27.79% | 0.331 | 0.383 | 18.85% | 0.679 | 0.371 | 14.96% | 0.727 | 0.353 | 14.82% | 0.739 |

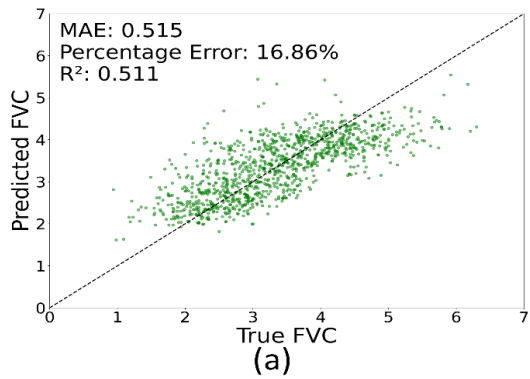

MAE: 0.515
Percentage Error: 16.86%
$R^2$: 0.511

(a)

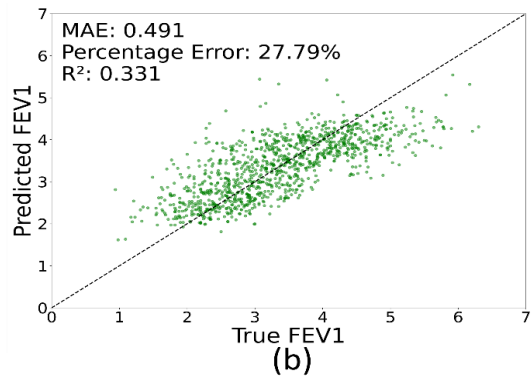

MAE: 0.491
Percentage Error: 27.79%
$R^2$: 0.331

(b)

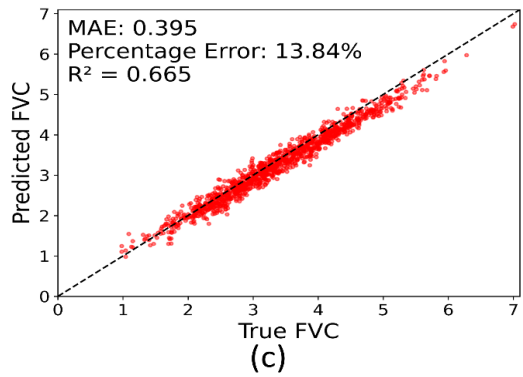

MAE: 0.395
Percentage Error: 13.84%
$R^2$ = 0.665

(c)

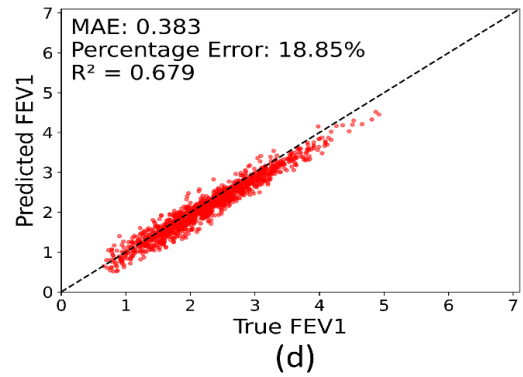

MAE: 0.383
Percentage Error: 18.85%
$R^2$ = 0.679

(d)

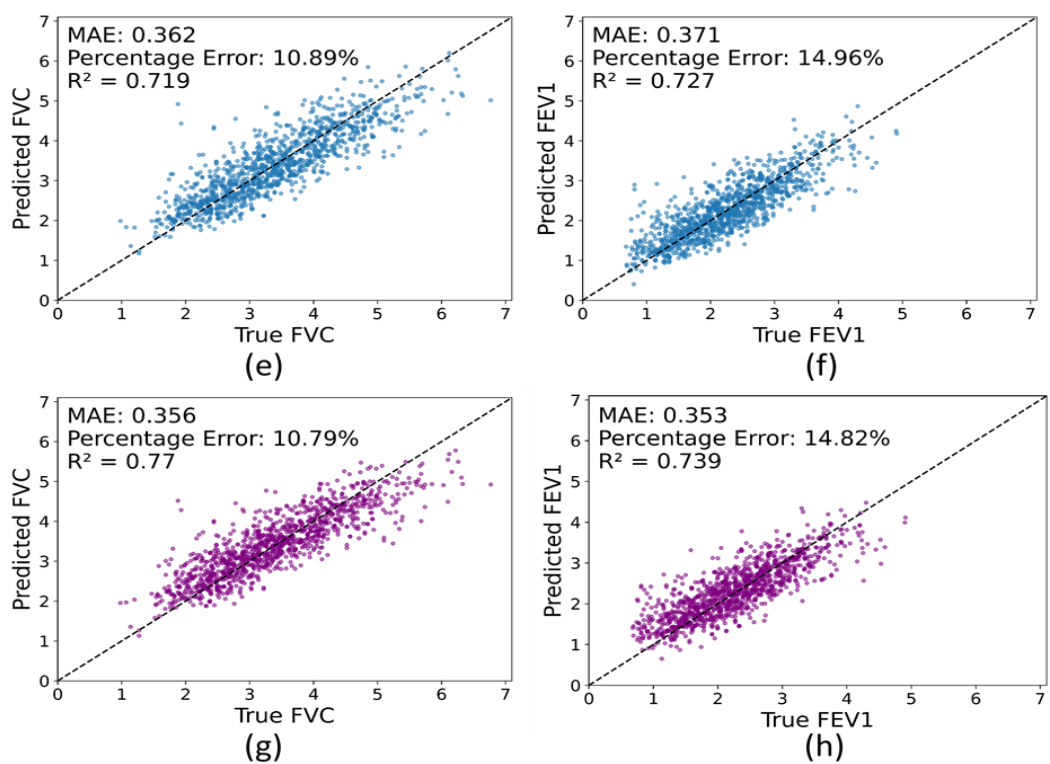

**Figure 2:** Scatterplots of the predicted versus observed FVC and FEV1 in the PLuSS independent test set (n = 362). (a) and (b) Linear regression using patient demographics. (c) and (d) 3D CNN. (f) BeyondCT without demographics.the (g) and (h) BeyondCT with demographics.

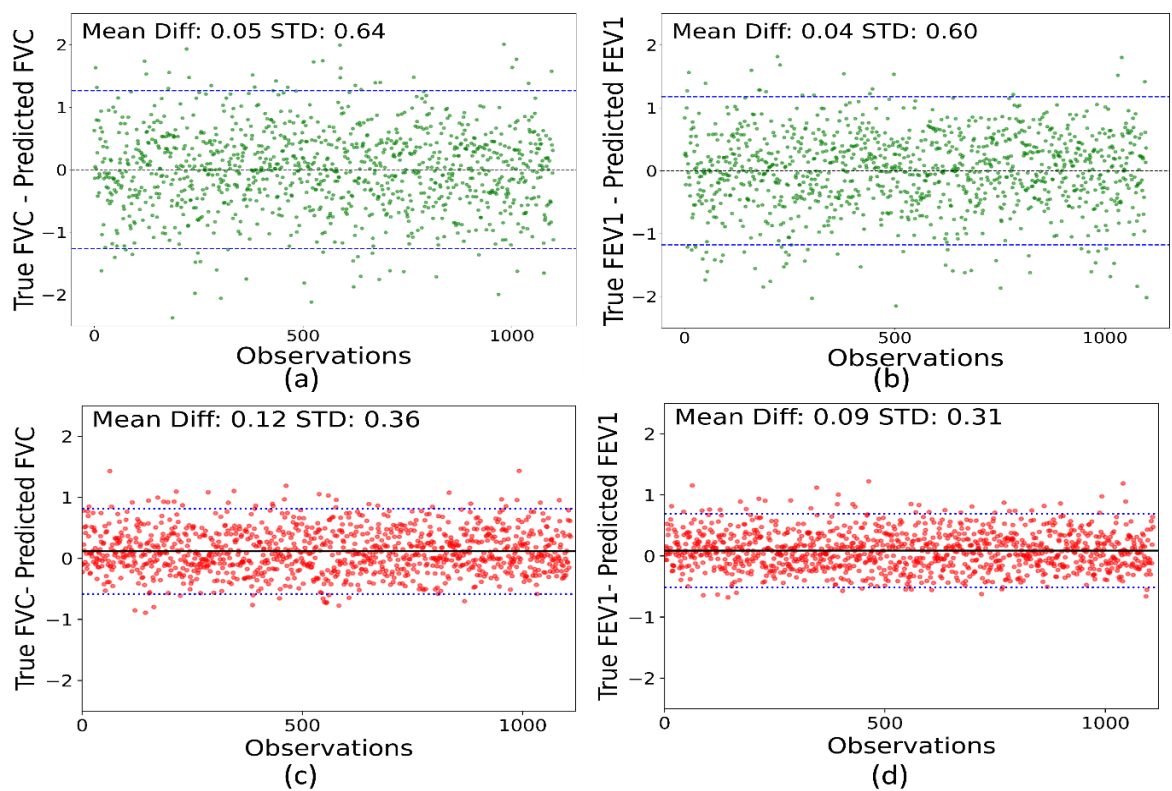

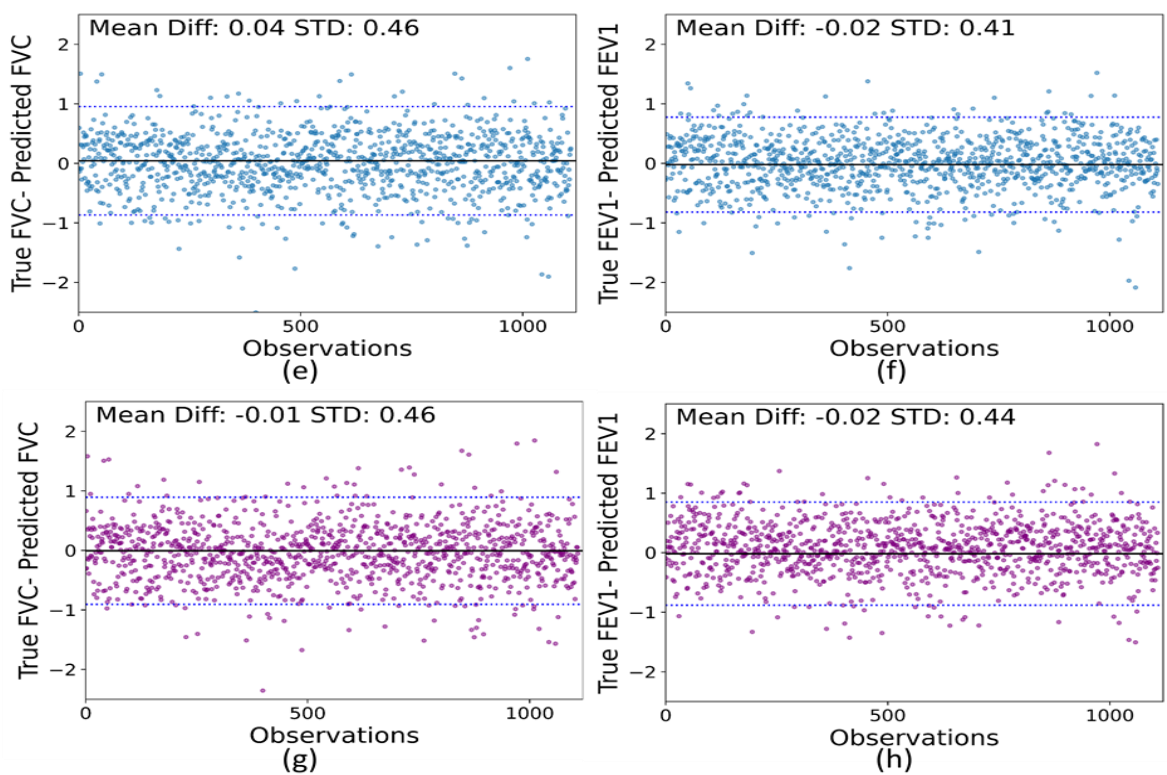

**Figure 3:** Bland–Altman plots of predicted versus observed FVC and FEV1 in the PLuSS independent test set. (a) and (b) Linear regression using patient demographics. (c) and (d) 3D CNN. (e) and (f) BeyondCT without demographics. (g) and (h) BeyondCT with subject demographics. The mean bias is indicated by the solid line, while the upper and lower limits of agreements are depicted by the blue dashed lines.

The BeyondCT model with demographics exhibited consistent performance based on gender, emphysema, GOLD criteria, and smoking status in the PLuSS independent test set (Table 4). In the emphysema group, the model had significantly smaller errors in the group without emphysema, but the difference was small.

**Table 4.** Performance of BeyondCT with demographics in subgroups of the PLuSS independent test set (n=362).

|  | Gender | Emphysema | GOLD Criteria | Smoking Status |
|---|---|---|---|---|
| **FVC %Error** | Male: 10.76% | Yes: 11.22% | 0:10.75% | Current: 10.76% |
|  | Female: 10.82% | No: 10.46% | 1-4: 10.83% | Former: 10.83% |
| **P value:** | 0.18 | 0.03 | 0.23 | 0.06 |
| **FEV1 %Error** | Male: 14.73% | Yes: 15.29% | 0: 14.78% | Current: 14.77% |
|  | Female: 14.91% | No: 14.35% | 1-4: 14.89% | Former: 14.86% |
| **P value:** | 0.07 | 0.04 | 0.15 | 0.42 |

## B. Performance on the SCCOR cohort

The BeyondCT model with subject demographics demonstrated significantly better prediction performance than the other models for both FEV1 and FVC (Table 5 and Fig. 4). The model had MAEs of 0.407 L and 0.412 L, percentage errors of 11.4% and 16.3% for predicting FVC and FEV1, respectively. The BeyondCT model with demographics had mean differences of 0.01 L and -0.02 L, lower limits of agreement of -1.6 L and -1.7 L, and upper limits of agreement of 1.8 L and 1.9 L for FVC and FEV1, respectively, in the Bland Altman analysis (Fig. 5).

**Table 5.** Models to predict FVC and FEV1 in the SCCOR independent test set (n=662)

|  | Linear regression with demographics | | Baseline CNN model | | BeyondCT without demographics | | BeyondCT with demographics | |
|---|---|---|---|---|---|---|---|---|
|  | MAE (L) | %Error | MAE (L) | %Error | MAE (L) | %Error | MAE (L) | %Error |
| FVC | 0.693 | 26.31% | 0.591 | 18.56% | 0.418 | 11.63% | 0.407 | 11.38% |
| FEV1 | 0.686 | 29.87% | 0.554 | 23.46% | 0.421 | 16.52% | 0.412 | 16.29% |

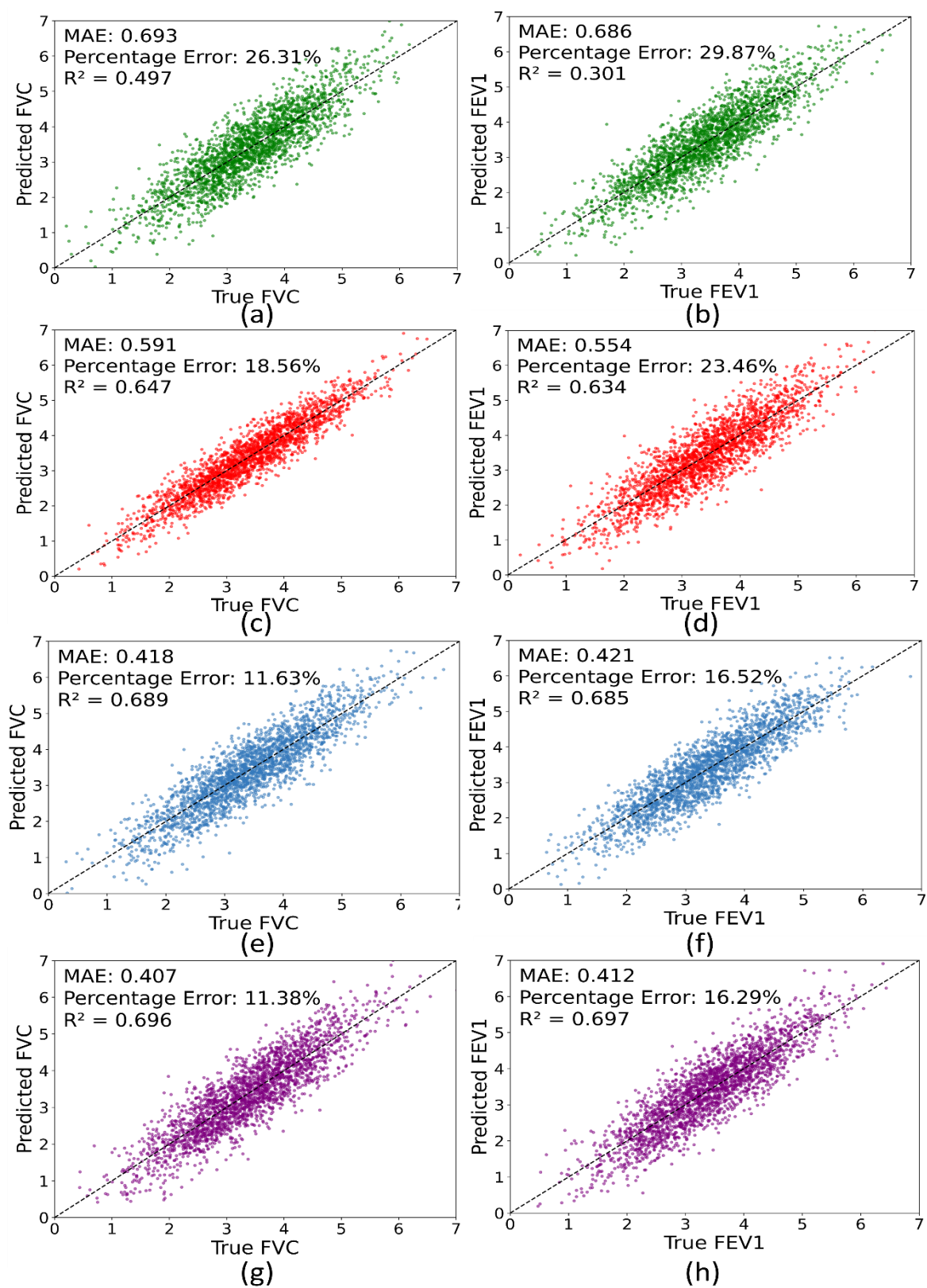

**Figure 4:** Scatterplots of the predicted versus observed FVC and FEV1 in the SCCOR test set (n = 662).
(a) and (b) Linear regression using patient demographics. (c) and (d) 3D CNN. (f) BeyondCT without
demographics. (g) and (h) BeyondCT with demographics.

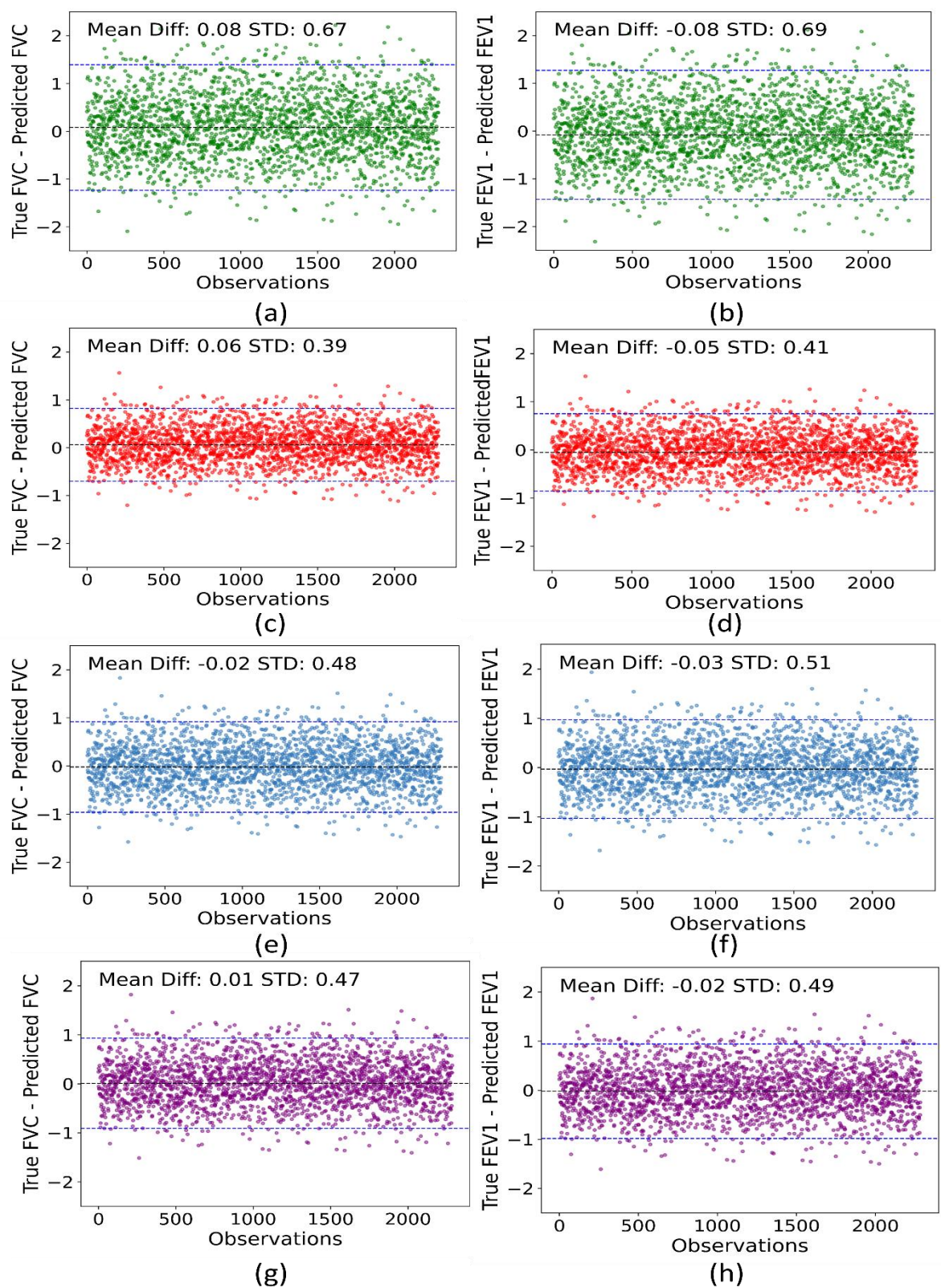

**Figure 5:** Bland–Altman plots of predicted versus observed FVC and FEV1 in the SCCOR test set. (a) and (b) Linear regression using patient demographics. (c) and (d) 3D CNN. (e) and (f) BeyondCT without demographics. (g) and (h) BeyondCT with subject demographics. The mean bias is indicated by the solid line, while the upper and lower limits of agreements are depicted by the blue dashed lines.

## C. Prediction errors on the two cohorts

The error rates of the BeyondCT with demographics model were essentially normally distributed, with approximately 50% of the samples having an error rate below the mean error for predicting both FVC and FEV1 (Fig. 6). The error rates ranged from 0-33% and from 0-35% in the PLuSS and SCCOR cohorts, respectively.

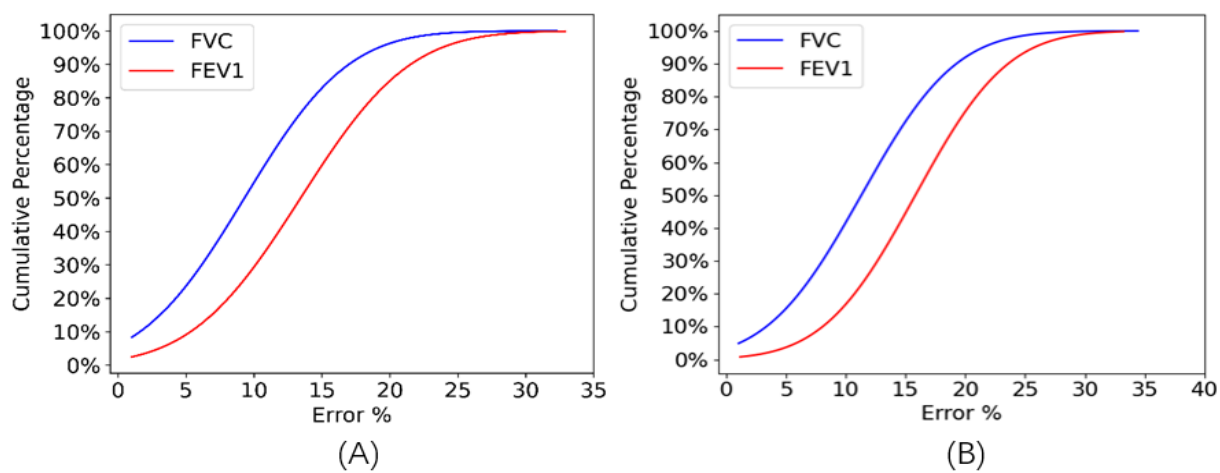

**Figure 6:** Cumulative distribution of the prediction error rate using BeyondCT with demographics on the PLuSS cohort (A) and the SCCOR cohort (B).

## D. GOLD classification

The BeyondCT with demographics model had a sensitivity of 92.5% and 95.3% and specificity of 95.7% and 94.1% to identify subjects with and without COPD based on GOLD criteria in the PLuSS and SCCOR independent test sets, respectively (Fig. 7). The model had a modest ability to identify the individual GOLD COPD severity categories for the PLuSS and SCCOR cohorts. The model's best performance to correctly identify subjects was in subjects with mild COPD (GOLD I) at 72.7% and 71.1% in the PLuSS and SCCOR cohorts, respectively.

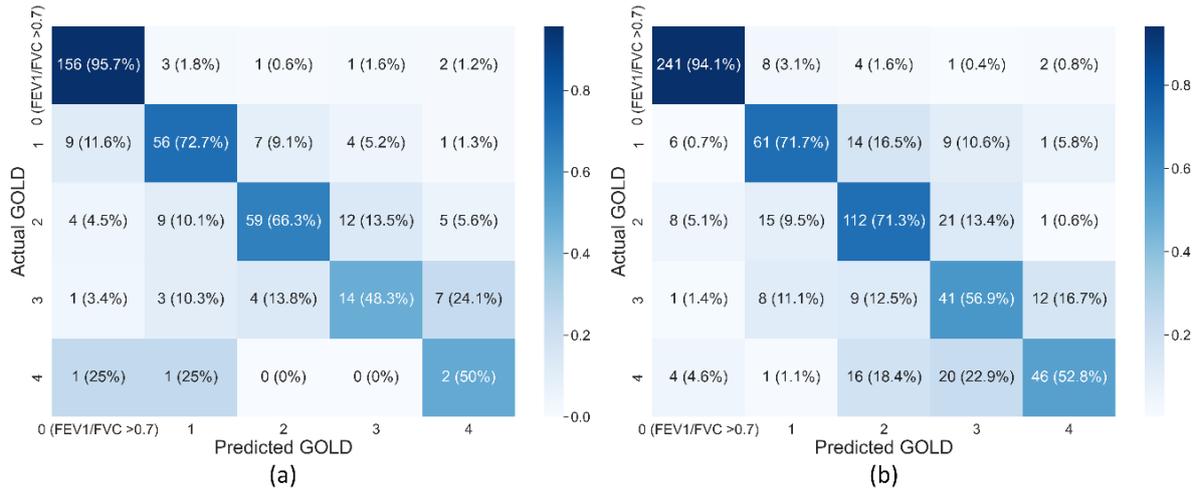

**Figure 7:** Confusion matrices to predict GOLD classifications based on the predicted FEV1/FVC values by the BeyondCT model with demographics on the PLuSS (a) and SCCOR (b) cohorts.

## IV.    Discussion

We developed and validated a novel deep learning model called BeyondCT to predict FVC and FEV1 values from chest CT scans and subject demographics. The model combines 3D CNN and ViT to accurately predict pulmonary function. While 3D CNNs can effectively extract spatial features from volumetric CT data, they may struggle with unusual or anomalous data points that may not be observable in local neighborhoods of pixels [11]. The ViT module in the BeyondCT model can utilize global information to determine whether an anomaly carries valuable information (such as an unusual image pattern indicating a rare occurrence of the disease process) or is merely noise [7]. Combining the local information extracted from the 3D CNN module and the global information derived from the ViT enables the BeyondCT model to achieve a comprehensive data understanding, leading to improved accuracy and robustness. BeyondCT performed significantly better than other modeling approaches (including a 3D CNN) to predict FVCX and FEV1 in two independent test sets from different cohorts (Tables 3 and 5). Our independent test sets originate from LDCT scans used to screen for lung cancer and a moderate dose CT scan to screen for COPD.

In the context of lung function prediction, it's important to consider the inherent variability associated with spirometry measurements. In standard clinical practice, spirometry tests are conducted at least three times to reliably determine lung functions. This practice stems from the dynamic nature of pulmonary tests, which can be influenced by patient effort, technique, and physiological variability. As pointed by Ferris et al. [12] and Miler et al. [13], acceptable repeatability in FVC measurements is achieved when the difference between the largest and the next largest FVC values is less than 0.150 L, underscoring the inherently dynamic nature of pulmonary tests. Given the inherent variability in FVC measurements, it's reasonable to expect that any model predicting lung function, including AI algorithms, would have unavoidable prediction errors. Therefore, our model's percentage error of a little over 10% is reasonable and promising.

BeyondCT predictions of FVC and FEV1 were extremely unbiased across the range of values (Figs. 3 and 5). The mean difference between the BeyondCT prediction and observed FVC and FEV1 value was consistently below 0.05 L (Fig. 3) The 3D CNN model predictions for FVC and FVC in the PLuSS subjects were systematically lower than the observed values with a mean differences of 0.12 L and 0.09 L, respectively (Figs. 3c and 3d), but not for the SCCOR subjects (Figs. 5c and 5d). Park et al. [3] observed an overestimation in the lower range and an underestimation in the higher range of FEV1 values by their CNN models compared to the observed values. They did not observe a systemic bias in predicting FVC and FEV1 using CNN models. The 3D CNN baseline model tended to have a higher bias and lower variance compared to BeyondCT with demographics for FVC and FEV 1 (Figs. 3 and 5). In contrast, BeyondCT demonstrated the opposite trend, with a lower bias and slightly higher variance for FVC and FEV1 predictions. The scatterplots in Fig. 2 showed that the 3D CNN model clustered data points closely, resulting in a narrower dispersion of predicted values but poor performance on outliers and thus a high bias. In contrast, the BeyondCT model (Fig. 2) exhibited a slightly broader dispersion of data points without apparent biases. The bias-variance tradeoff can be attributed to the increased complexity of the BeyondCT model, as previously discussed in [14].

BeyondCT demonstrated improved generalization and is less prone to overfitting compared to 3D CNN models. The 3D CNN model exhibited a notable increase in MAE and percentage error when switching from the training set to the test sets, suggesting that the model's performance deteriorated as it attempted to generalize to unseen data and possible overfitting (Table 3). This observation may explain the lack of replicability of the model proposed by Park et al. [3] on our dataset. This discrepancy underscores the importance of model robustness and generalization in clinical machine learning applications, for which our BeyondCT model has demonstrated its significance. Its relative consistency across the training set, independent test set, and the SCCOR cohort (Table 3, 5) implies better generalization and less likelihood of overfitting compared to the 3D CNN model.

The inclusion of demographics significantly improved in the prediction accuracy of BeyondCT for both FVC and FEV1 in both the PLuSS and SCCOR datasets (Tables 3 and 5). Significant reductions in error rates were observed in most cases when demographic data was included, with only a slightly smaller difference observed in FVC prediction in the PLuSS validation set. For both FVC and FEV1 prediction, of the BeyondCT demographics model had a higher $R^2$ values for both FVC and FEV1 in the PLuSS training, validation, and test datasets (Table 3). One explanation for this improvement is that demographic variables carry essential contextual information that influence a patient's general pulmonary health condition beyond anatomical features not captured in CT scans [15-17]. Therefore, the multimodal incorporation of demographics with anatomical characteristics captured by CT images may provide a more comprehensive understanding of the individual's physiological condition than either assessment alone. By fusing both CT scans and patient demographics into a unified input, the model achieved better generalizability and robustness to data quality variations. When CT image data is incomplete or of low quality, the information from the demographics can compensate and help maintain the model's performance.

We investigated embedding single and subsets of demographic features, but no significant differences were observed compared to the BeyondCT without patient demographics. However, statistically significant differences were observed in the prediction of FVC (p-value: 0.03) and FEV1 (p-value: 0.04) between subjects with and without emphysema (Table 4). The presence of emphysema presented additional challenges for accurately predicting FVC and FEV1 from CT scans because of the presence of bullae, as it led to the destruction of lung air sacs and the formation of larger but fewer sacs. While the inclusion of demographics for improved prediction is straightforward, our study aims to clarify the contributions of the demographic factors comprehensively.

The automated BeyondCT model predicts lung function directly from CT scans, which has several advantages to other approaches in the literature. Chen et al.'s methodology requires features extraction and collecting inflammatory parameters, which includes tedious labeling and feature extraction [1]. Wang et al.'s approach requires FEV1 as an input to predict FVC, which limits its application to cases with existing PFT data[2]. Similarly, Javaregowda et al. approached heavily relied on historical FVC measurements [4]. Hence, BeyondCT is more clinically practical compared to these methods.

This study has several limitations. First, the models were trained exclusively on the PLuSS cohort acquired from a single institution. However, the PLuSS CT scans were acquired over a period of time using different scanners and protocols, which resulted in a range of image quality. To reduce this limitation, the SCCOR cohort acquired using different protocol was used to independently validate the performance of the models, which demonstrated BeyondCT's robust performance. This could be attributed to the data augmentation operation that might alleviate the differences of the image qualities across the two cohorts. Second, the subjects in the cohort are smokers aged over 50, which may affect the model's generalization to the general or different populations. However, this needs to be verified in a cohort representing the general population including smokers and non-smokers. Third, the "black box" nature of deep learning models presents challenges in terms of interpretability. Although our model demonstrates promising prediction performance, its inability to provide explicit explanations for its decision-making process may restrict its clinical applications. For a deep learning based regression task, available technologies such as Grad-CAM and NUN as demonstrated in our previous studies [18, 19], are not applicable as they were designed for classification only. Additionally, unlike tumor detection or classification, the factors affecting lung functions, such as the presence of lung disease, body composition, and physical fitness, are heterogeneous and vary significantly across the subjects. As a result, they do not follow certain patterns. Hence, it is very difficult to visualize the regions of interest based on the attention or activation map that can help explain the regression results. Nevertheless, additional investigative effort is needed to explain the results of a regression model.

## V. Conclusion

We presented a deep learning framework called BeyondCT that combines 3D CNN and Vision Transformer architectures to accurately predict pulmonary function from CT scans.

BeyondCT outperforms baseline 3D CNN models, demonstrating improved accuracy, robustness, and generalizability. The inclusion of demographics improved prediction accuracy., and the model exhibits resilience to outliers. However, limitations include the lack of dataset diversity and interpretability challenges. BeyondCT represents a significant advancement in pulmonary function prediction and holds promise for improving clinical diagnosis and management of lung diseases.

**Acknowledgment**

This work is supported in part by research grants from the National Institutes of Health (NIH) (R01CA237277, U01CA271888, and R61AT012282)

**Reference:**

1.  Chen, J., et al., *Prediction models for pulmonary function during acute exacerbation of chronic obstructive pulmonary disease.* Physiol Meas, 2021. **41**(12): p. 125010.

2.  Wang, C., et al., *Predicting forced vital capacity (FVC) using support vector regression (SVR).* Physiol Meas, 2019. **40**(2): p. 025010.

3.  Park, H., et al., *Deep Learning-based Approach to Predict Pulmonary Function at Chest CT.* Radiology, 2023. **307**(2): p. e221488.

4.  Javaregowda, M., et al., *Prediction of Pulmonary Fibrosis Progression using CNN and Regression*. 2021. 944-950.

5.  Schroeder, J.D., et al., *Prediction of Obstructive Lung Disease from Chest Radiographs via Deep Learning Trained on Pulmonary Function Data.* Int J Chron Obstruct Pulmon Dis, 2020. **15**: p. 3455-3466.

6.  Tran, D., et al. *Learning Spatiotemporal Features with 3D Convolutional Networks*. in *2015 IEEE International Conference on Computer Vision (ICCV)*. 2015.

7.  Dosovitskiy, A., et al., *An image is worth 16x16 words: Transformers for image recognition at scale.* arXiv preprint arXiv:2010.11929, 2020.

8.  Vaswani, A., et al., *Attention is all you need.* Advances in neural information processing systems, 2017. **30**.

9.  Shi Z, Geng K, Zhao X, et al. XRayWizard: Reconstructing 3-D lung surfaces from a single 2-D chest x-ray image via Vision Transformer. Med Phys. 2023; 1-11. https://doi.org/10.1002/mp.16781

10. Kingma, D.P. and J. Ba, *Adam: A method for stochastic optimization.* arXiv preprint arXiv:1412.6980, 2014.

11. Yamashita, R., et al., *Convolutional neural networks: an overview and application in radiology.* Insights Imaging, 2018. **9**(4): p. 611-629.

12. Ferris, B.G., Jr., et al., *Spirometry for an epidemiologic study: deriving optimum summary statistics for each subject.* Bull Eur Physiopathol Respir, 1978. **14**(2): p. 145-66.

13. Miller, M.R., et al., *Standardisation of spirometry.* Eur Respir J, 2005. **26**(2): p. 319-38.

14. Yang, Z., et al. *Rethinking bias-variance trade-off for generalization of neural networks*. in *International Conference on Machine Learning*. 2020. PMLR.

15. Chuang, M.-L., I.-F. Lin, and K. Wasserman, *The body weight–walking distance product as related to lung function, anaerobic threshold and peakV̇O2in COPD patients.* Respiratory medicine, 2001. **95**(7): p. 618-626.

16. Gunnell, D., et al., *Associations of height, leg length, and lung function with cardiovascular risk factors in the Midspan Family Study.* Journal of Epidemiology & Community Health, 2003. **57**(2): p. 141-146.

17. Wahba, W., *Influence of aging on lung function-clinical significance of changes from age twenty.* Anesthesia & Analgesia, 1983. **62**(8): p. 764-776.

18. Singh, J., et al., *Batch-balanced focal loss: a hybrid solution to class imbalance in deep learning.* J Med Imaging (Bellingham), 2023. **10**(5): p. 051809.

19. Beeche, C., et al., *Assessing retinal vein occlusion based on color fundus photographs using neural understanding network (NUN).* Med Phys, 2023. **50**(1): p. 449-464.